\documentclass[11pt,prb]{revtex4}

\usepackage{amsmath}  % need for subequations
\usepackage{graphicx}

\begin{document}

%------------------------------------------------------------TITLE +  AUTHORS
\title{Flux flow properties of superconducting niobium thin films in clean and dirty limits }
\author{Christophe Peroz and Catherine Villard}
\address{Centre de Recherches sur les Tr\`es Basses Temp\'eratures\\             
Consortium de Recherches pour l'Emergence de Technologies Avanc\'ees\\ 
CNRS, 25 avenue des Martyrs, 38042 Grenoble, France-Communaut\'e Europ\'eenne} 

%------------------------------------------------------------ABSTRACT
\begin{abstract}
Flux flow properties for clean and dirty superconducting limits in strong pinning niobium films are studied. Measurements of electric field vs current density characteristics at high $J$ values ($J\simeq 10^{6} A/cm^{2}$) are successfully compared to theoretical models of flux flow and of flux flow instability in a large range of temperatures and magnetic fields. The non-linear regime at high dissipation is analysed in the frame of a modified Larkin Ovchinnikov model that takes into account a quasiparticles  heating effect. This model elaborated by Bezuglyj and Shklovskij (BS) defines a transition magnetic field $B_{t}$ above which the quasiparticles  distribution became non uniform due to a finite heat removal from the substrate. From the BS model, we can deduce values of the non-equilibrium lifetime of quasiparticles  $\tau_{qp}$, which are 10 to 100 times shorter in the dirty sample compared to the clean one and whose temperature dependance is specific to the electronic nature of the Nb film. The study of the non-linear regime provides also a quantitative determination of the thermal transparency of the film-substrate interface. 

\end{abstract}

\maketitle

%------------------------------------------------------------TEXT

\section{Introduction}
Magnetic vortices in the mixed state of a superconductor can move under the action of the Lorentz driving force $\overrightarrow{F_{L}}$ proportional to a transport current density $J\!-\!J_{c}$, where $J_{c}$ is the pinning threshold. Vortex dynamics near $J_{c}$ has been recently widely explored by direct methods such as magneto-optics\cite{johansen}. Inversely, the regime at high vortex velocities $v_{v}$ is only deduced  from electric field vs current density characteristics $E(J)$ in accordance with the Josephson\cite{Josephson} equation $\overrightarrow{E}=-\overrightarrow{v_{v}}\times \overrightarrow{B_{a}}$, where $\overrightarrow{B_{a}}$ is the magnetic field inside the superconductor. A first experiment\cite{Kim64} in 1964 has interpreted a linear dependence between $E$ and $J$ as a steady flux flow (FF) of vortices. The FF regime is defined by the balance between $F_{L}$
%, opposed to a pinning force $\overrightarrow{F_{p} } = \overrightarrow{J_{c}(v,B)} \times \overrightarrow{B_{a}}$, 
and a viscous drag force $\overrightarrow{F_{\eta}}=-\eta_{FF}\overrightarrow{v_{v}}$ where $\eta_{FF}$ is the viscosity associated to vortex motion. A linear regime $J=\sigma_{FF}E+J_{c}$ for which the viscous coefficient $\eta_{FF}$ and the conductivity $\sigma_{FF}$ ($\sigma_{FF} = \frac{\eta_{FF}}{B.\Phi_{0}}$) are constant is thus expected. The dissipation in a moderate ``clean'' superconductor at low magnetic fields and temperatures is essentially due to currents crossing vortex cores\cite{BS65,Rosen04} : 
%-----------------------------------------------------------EQUATION DANS SYSTEME PROPRE------
 \begin{equation}\frac{   \sigma_{FF}   }{ \sigma_{n}(T=0)  } \simeq  \frac{  B_{c2}(T)   }{ B_{a}  }  - k(T),  \end{equation} with $\sigma_{n}(T=0)$ the conductivity of the normal state, $B_{c2}(T)=\frac{\Phi_{0}}{2\pi \xi^{2}}$ and $k(T)$ a temperature dependent positive constant representing the dissipation due to Cooper pairs around vortex cores. This equation can be simplified in:
\begin{equation}\frac{   \sigma_{FF}   }{ \sigma_{n}(T=0)  } =A(T)/ B_{a} - k(T)  \end{equation} where $A(T)$ is a temperature dependent coefficient.
%-----------------------------------------------------------
The expression of the conductivity $\sigma_{FF}$ for a ``dirty'' system adds the contribution of thermal effects\cite{Clem68} in the vicinity of the vortex core :
%-----------------------------------------------------------EQUATION DANS SYSTEME SALE------
 \begin{equation} \frac{   \sigma_{FF}   }{ \sigma_{n} (0)  } \simeq  \frac{ B_{c2}(T)+\Im (t)\times B_{c2}(0) }{ B_{a}  } -k(T), \end{equation} where $\Im (t)$ is a positive function with a maxima at $t=T/T_{c}\simeq 0.5$.\\
%-----------------------------------------------------------
At higher velocities $v_{v}$, Larkin and Ovchinnikov\cite{LO75} (LO) have calculated a non-linearity of the conductivity $\sigma_{FF}(v_{v})$ for temperatures $T$ close to $T_{c}$ and $B_{a}\ll B_{c2}$. High electric fields in the vicinity of the vortex core can change the electronic distribution and the electronic temperature $T_{qp}$ leading to a decrease of $\eta_{FF}$ when $v_{v}$ increases. $E(J)$ characteristics become non-linear and end with a jump into the normal state before the depairing current density is reached. In the absence of thermal runaway, the flux flow instability occurs at a critical vortex velocity $v_{v}^{*}$. To come to this critical point ($J^{*}, E^{*}$), the quasiparticles  inside the driven vortex cores have gained enough energy from the electric field to escape from these normal regions and relax their energy into the condensate. This process is controlled  by a non-equilibrium lifetime $\tau_{qp}$ of the quasiparticles. This electronic leakage leads to a continuous shrinkage of the vortex core radius $\xi$ and to a decrease of $\sigma_{FF}$ as a function of $v_{v}$.

In the original LO theory, the non linear flux flow behavior is due only to the field induced change in the quasiparticle distribution function and not Joule heating: the system is in thermal equilibrium with the bath, characterized by a temperature $T_{0}$. Several experimental works in low\cite{Mu80,Kl85,Per02} and high\cite{Doe94,Xiao96} $T_{c}$ superconductors have confirmed the validity of the LO model near the critical temperature, with however some discrepancies. \\
It comes out that the sample heating during the dissipative flux flow is not negligible and can yields a thermal runaway\cite{Gonzales03} before the occurrence of the FF instability. Bezuglyj and Shklovskij\cite{BS92} (BS) have extended the LO theory in the thin film configuration by taking into account heating effects.  In their model, the quasiparticles  distribution function depends on the vortex density and on the rate of heat removal from the film through the substrate. BS defines a transition magnetic field $B_{t}$, dependent on the thermal exchange coefficient $h$ between the film and the substrate, under which the hypothesis of a uniform quasiparticles  distribution (LO model) is still valid : 
%-----------------------------------------------------------DEFINITION DE B_{t}-----
 \begin{equation} B_{t}=\frac{0.374\,e\,h\,\tau_{qp}        }{  k_{B}\,\sigma_{n}\,d    }, \end{equation} where $e$ is the electronic charge, $k_{B}$ the Boltzman constant and $d$ the film thickness.
%-----------------------------------------------------------
For $B_{a}>\!>B_{t}$, dissipation during the FF regime raises the electronic temperature $T_{qp}$ and thermal effects govern the FF instability. This effect can be understood from a simple dynamical picture where the inter-vortex spacing becomes small enough to allow an influence of a vortex core on the other. In other words, the condensate keeps a memory in terms of quasiparticles  energy (and temperature) from vortex passing.  In contrast to the B-independent $v_{v}^{*}$ of the  LO model, a $v_{v}^{*} (B_{a})$ variation is now expected\cite{BS92} and takes the form : 
%-----------------------------------------------------------DEPENDANCE MAGNTQ DE  v*-----
 \begin{equation}  v_{v}^{*} \propto  h\;(1-t)^{1/4}B_{a}^{n}, 
  \end{equation} 
 %-----------------------------------------------------------
with $n=-0.5$. BS proposed a scaling law between critical parameters $J^{*}$ and $E^{*}$ for the heated quasiparticles  at $T_{qp}>\!>T_{0}$ :
%-----------------------------------------------------------COURBE UNIVERSELLE-----
 \begin{equation} \frac{E^{*}     }{  E^{*}_{LO}  } =  (1-z(b_{t}))(\frac{J^{*}     }{  J^{*}_{LO}  })^{-1}, 
 \label{eq6} \end{equation} 
%-----------------------------------------------------------
where $E^{*}_{LO}$ and $J^{*}_{LO}$ are critical parameters of the pure magnetic (LO) theory and $z(b_{t})$ is a function of $b_{t}=B_{a}/B_{t}$.\\
Recent works\cite{Kun02, Babic04} at low temperatures ($T \leq 0.4T_{c}$) suggest that thermal effects can diminish the superconducting order parameter and lead to an expansion of vortex cores rather than to a shrinkage. The quasiparticles  heating reduces critical field $B_{c_{2}}(T)$ to $B_{c_{2}}(T^{*}_{qp})=B_{a}$ where the transition to the normal state occurs above $v_{v}^{*}$. The $B_{a}$ dependance of $v_{v}^{*}$ is the same than the one given by the BS theory.\\
In this paper, we explore and compare the dynamic of a vortex lattice at high velocities in niobium micro-bridges for ``clean'' and ``dirty'' superconducting limits in a range of temperature above $0.6T_{c}$.  We reveal the influence of the electronic nature of superconductors on flux flow properties in the framework of the BS model. Values of $h$ and the temperature behaviour of $\tau_{qp}$, which are  important intrinsic parameters for the development of electronic devices such as hot electrons bolometers, are also deduced.
%
%
%-----------------------------------------------------------EXPERIMENTAL DETAILS
%
%
\section{Experimental details}

Niobium thin films with thickness $d\simeq100 nm$ are prepared by ion beam technique. Depositions are done either at ambient temperature (cool sample in dirty limit) or at 780\r{}C (warm sample in clean limit) on  Si and Al$_{2}$O$_{3}$  substrates. Films are protected by a silicon thin layer of $5nm$ thickness. Microbridges  of $8.5$ to $10\mu m$ width ($w$) are patterned to achieve a four points configuration measurement. Bridge lengths between voltage contacts are included between $l=800\mu m$ and $l=3 mm$. 
 Pinning is strong in niobium films: values of critical current densities $J_{c}$ are typically a few $10^{6}A/cm^{2}$ at $0.7T_{c}$. These sample parameters are within the range of expected values for Nb films. In all the experiments presented here, the magnetic field is applied perpendicular to the samples surface. More details are given in reference 17. 
%----------------------------------------------------------------TABLE
\begin{table}[tbp]
\caption{Example of fundamental characteristics for ``dirty'' and ``clean'' Nb films}
\label{t.1}
\begin{center}
\begin{tabular}{l ||c| c| c| c| c| c| c|  }
Sample & substrate & $T_{c_{H=0}}$ (K) & $H_{c2}$(0) (mT) & $\rho _{n_{9.2K}}$ ($\mu
\Omega $.cm) & $\xi _{0_{BCS}}$ (nm) & $l_{free}$ (nm)  \\ 
\hline \hline
Nb dirty & $Al_{2}O_{3}$ &8.92 & 4600 & 12.12 & 36  & 3.19 \\ 
Nb clean & $Al_{2}O_{3}$ & 9.13 & 1010 & 0.59 & 33 & 65 \\
Nb dirty & $Si$ & 8.6 & 4430&9.9 & 35 & 3.9  
\end{tabular}
\end{center}
\end{table}
%----------------------------------------------------------------

%
%				FIGURE 1
\begin{figure}[!h]
\begin{center}
\includegraphics[width=9cm]{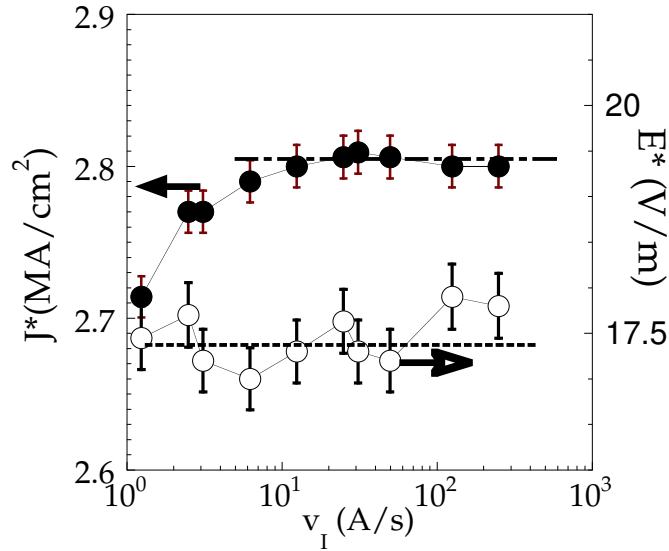}
\caption{\label{figure1}  $E(J)$ curves at $T=7.8K$ ($t\simeq 0.9$) for $0$ $<$ $B_{a}$ $<$ $150 mT$  ($B_{c2}(T) =$ $420 \pm 8$ $mT$ ) in a ``dirty'' niobium film.
}
\end{center}
\end{figure}
%-------------------------------------------------------------

\begin{figure}[!h]
\begin{center}
\includegraphics[width=8cm]{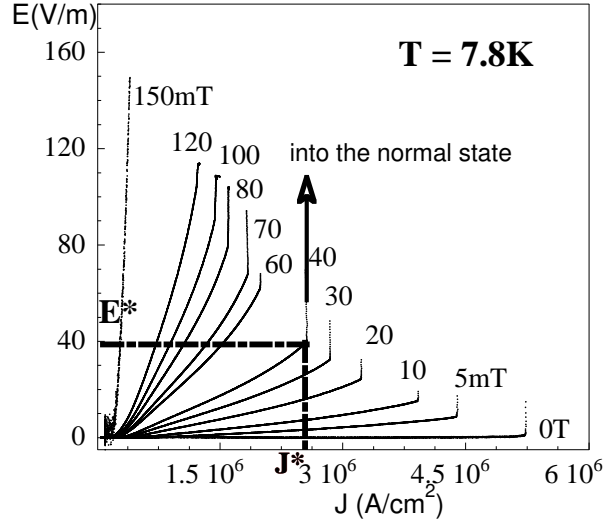}
\caption{\label{figure2} Variations of $J^{*}$  and $E^{*}$ vs $v_{I}$ at $T=6K$ and $B_{a}=20mT$ in a ``dirty'' niobium film. See Fig. 1 for the experimental determination of  $J^{*}$  and $E^{*}$. These critical parameters are independent of the sweep rate above 20A/s.
}
\end{center}
\end{figure}

Current-voltage characteristics were measured through fast current sweeps ($v_{I}\in [1;250A/s]$). An example is reported in Figure \ref{figure1}. The electric field $E$ and the current density $J$ were determined from relations $E=V/l$ and $J=I/S$ where $l$ and $S=d\times w$ are respectively length and section of microbridges. Parasitic thermal effects coming from contact resistance and classical Joule dissipation due to vortex motion are identified by performing experiments at different current sweep rates. To retain only non-equilibrium effects pertinent for the BS model,  a high enough current  rate $v_{I}$ where the parameters $J^{*}$  and $E^{*}$ are constant (see figure \ref{figure2}) is applied.
%
%				FIGURE 1
%-------------------------------------------------------------
%
%---				RESULTS & DISCUSSIONS
%
\section{Results and discussion}
Figure  \ref{figure1} depicts a serie of typical $(E,J)$ curves in a ``dirty'' niobium sample at $t\simeq 0.9$ for $0$ $<$ $b(t)=B_{a}/B_{c2}(T)$ $<$ $0.35$. For current densities $J>J_{c}$, a linear regime $E\propto J$ occurs, which corresponds to a constant viscosity $\eta_{FF}$ (see also Fig. 3a). 
At higher currents, the response becomes non-linear (decreasing $\eta_{FF}$) and finally ends by a flux flow jump at low magnetic fields. This general behavior is reported for  both kinds of superconducting films.\\
The flux flow can be characterised by two thresholds of vortex speed, one defining the onset ($v_{min}$) of the linear FF regime, the other its end ($v_{max}$). It is interesting to note that these FF velocity thresholds are independent of the applied field, at least in the intermediate field range 10mT-70mT as shown in Fig. 3b. A similar behaviour is found in the clean limit film with however much lower $v_{max}$ (about 110 m/s).

%          FIGURE 3
\begin{figure}[!h]
\begin{center}
\includegraphics[width=14cm]{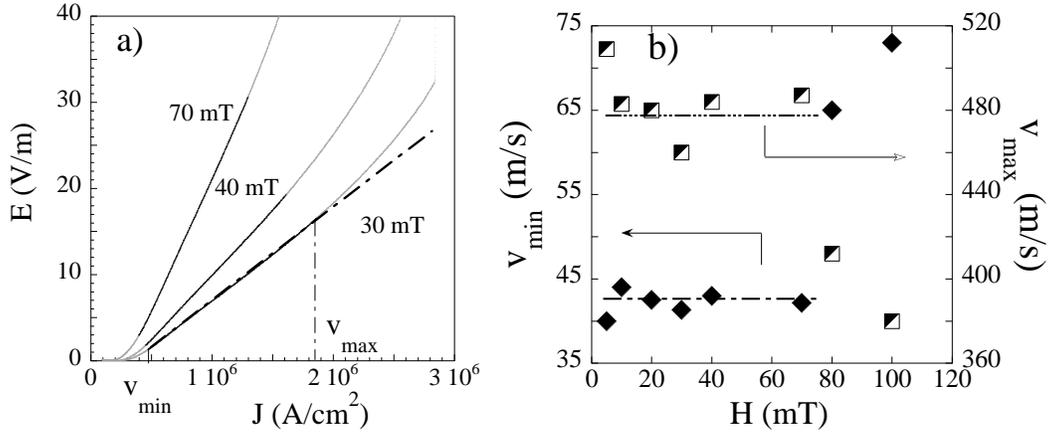}
\caption{\label{figure3} Right : $E(J)$ curves at $T=7.8K$ ($t\simeq 0.9$) for $B_{a} = 30,40, 70 mT$  in a ``dirty'' niobium film. Left : Magnetic field dependence of the thresholds of the vortex velocity $v_{min}$ and $v_{max}$.}
\end{center}
\end{figure}
\begin{figure}[!h]
\begin{center}
\includegraphics[width=9cm]{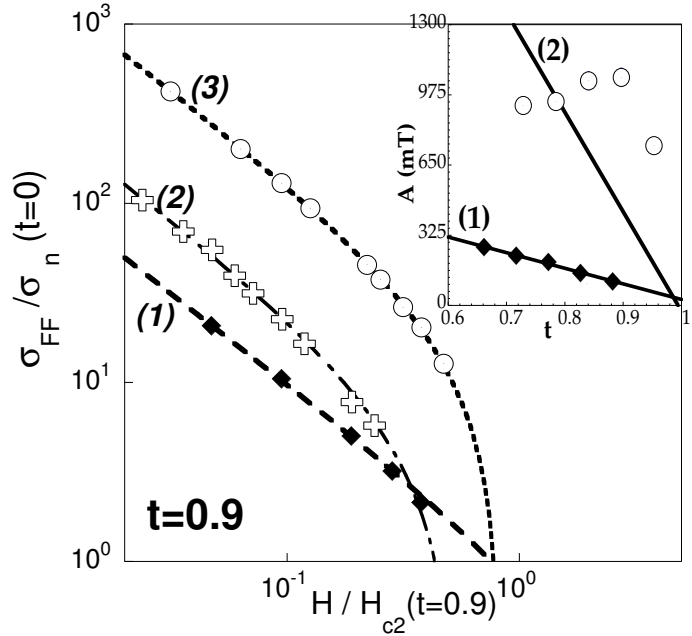}
\caption{\label{figure4} Ratio $\sigma_{FF}/\sigma_{n}$ as function of $b(t)$ at $t\simeq 0.9$. Sample (1) and (2) are respectively ``clean'' and ``dirty'' films on saphire substrate, and sample (3) is a  ``dirty'' film on a Si substrate. Dashed lines represent fits to equation $A(t) / B_{a} -k(t)$ with $A(t)$ and $k(t)$ as fitting parameters. Inset : Comparison of $A(t)$ points and  $H_{c2}(t)$ curves (solid lines). For sample $(1)$, $H_{c2}(t)$ is corrected by a factor $1.035$.
}
\end{center}
\end{figure}
\begin{figure}[!h]
\begin{center}
\includegraphics[width=9cm]{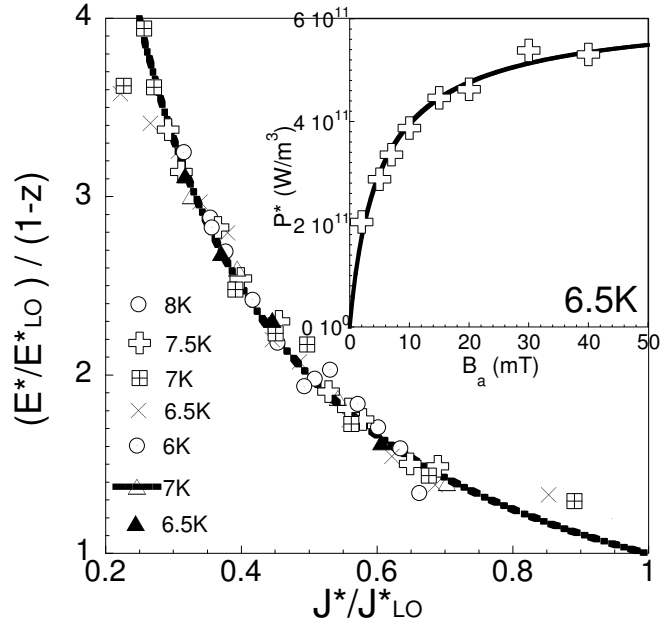}
\caption{\label{figure5} Comparison between $(E^{*},J^{*})$ data and universal scaling law $y=1/x$. Open symbols are for the  ``dirty'' film at $T=6;6.5;7;7.5;8K$. Triangle symbols are for the ``clean'' sample at $T=6.5;7K$. Inset : $P^{*}=P^{*}_{LO}[1-z(\frac{B_a}{B_t})]$ variation as a function of $B_{a}$ at $T=6.5K$ for the  ``dirty''  film. The two parameters $P^{*}_{LO}$ and $B_t$ are deduced from this fit. }
\end{center}
\end{figure}
\begin{figure}[!h]
\begin{center}
\includegraphics[width=9cm]{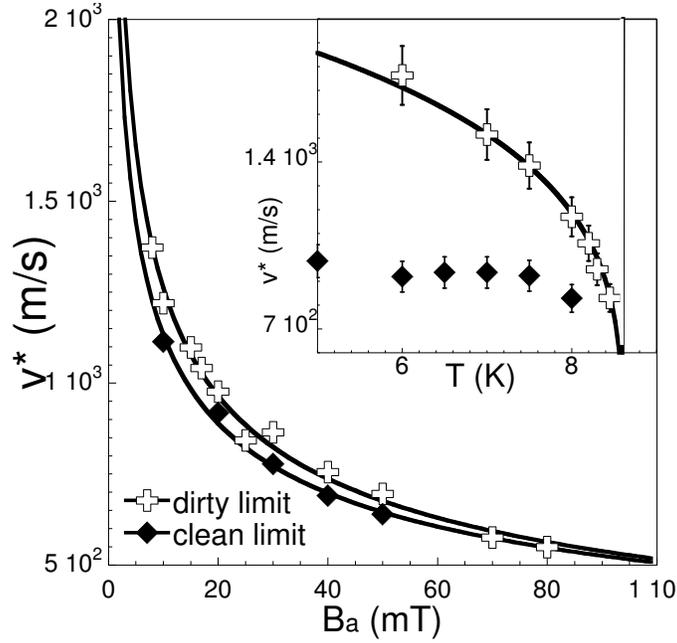}
\caption{\label{figure6} Extracted $v^{*}(B_{a})$ values for ``clean'' and  ``dirty'  films at $T=7.5K$. Solid lines fit data with $B_{a}^{n}$ : $n\simeq 0.35$ and $n\simeq 0.38$ respectively for ``clean'' and ``dirty'' film. Inset : corresponding $v^{*}(T)$ for $B_{a}=20mT$. Solid line represents $(1-T/T_{c})^{0.25}$ fit curve.
}
\end{center}
\end{figure}
\begin{figure}[!h]
\begin{center}
\includegraphics[width=7cm]{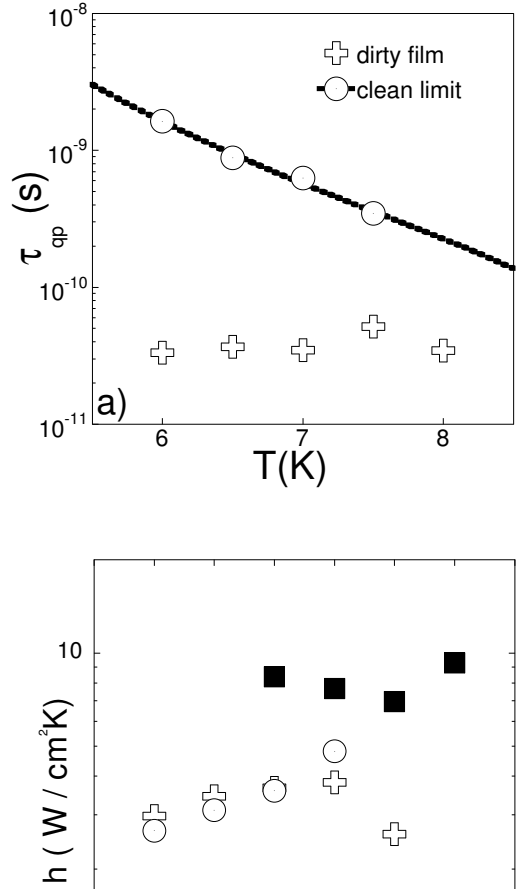}
\caption{\label{figure7} 
Variations of duration $\tau_{qp}(T)$ (Fig. a) and coefficient $h(T)$ (Fig. b). Dotted line in Figure $a$ fits experimental data with recombination model $e^{2\Delta (T)/k_{B}T}$. Squares symbols are data for $Au/Nb/Au$ on saphire substrate.
}
\end{center}
\end{figure}

The constant value of the onset of the linear regime can be interpreted by a dynamic phase transition\cite{Koshelev94,KOK04} under current from the motion of the amorphous vortex configuration at $J<J_{c}(v=v_{min})$ (plastic flow) to the motion of the vortex crystal (moving crystal) at $J>J_{c}(v=v_{min})$. The crystalliszation current is expected to exceed depinning current $J_{dep}$ and is related to lattice defect concentration. Alternatively,  a transition from vortex creep to flux flow with increasing Lorentz force can be considered. The existence and the value of the maximum vortex speed ending the linear FF regime will be discussed below in the section dedicated to the non-linear behaviour at high dissipation. \\   
The conductivity $\sigma_{FF}$ in the linear regime is treated within the ranges  $0.02<b<0.5$ and $0.65<t<0.95$ in the framework of the flux flow model. Figure \ref{figure4} analyses the $B_{a}$ dependence of the ratio $\sigma_{FF}/\sigma_{n}$ and shows the good accordance of experimental data with the $A(T) / B_{a} -k(T)$ formula for both kind of samples. For ``clean'' systems, data on almost two decades of dissipation agree with equation $(2)$, giving $A(T)$ absolute values shifted by only $3.5\%$ from the $H_{c_{2}}(T)$ curve deduced from  $AC$ resistivity measurements (see inset of figure \ref{figure4}). This result validates the approximation of a normal core with a radius of the order of $\xi(T)$ inside which the dissipation occurs. On the contrary, the coefficient $A(T)$ for ``dirty'' films doesn't not follow the temperature variation given by equation (2) and displays a maximum at $t\simeq 0.85$. For temperatures close to $T_{c}$, $A(T)$ decreases toward the normal conductivity but remains above the $B_{c_2}$ values, meaning that the ratio $\sigma_{FF}/\sigma_{N}$ is higher than expected, i.e. the dissipation inside the core is lower than what is found in clean samples. Larkin and Ovchinnikov\cite{LO86} have calculated  $\sigma_{FF}$ in the theoretical frame of the Ginzburg-Landau theory. Their formula depending on $T$ and $B_{a}$ doesn't fit our present data for ``dirty'' samples. A complementary theoretical work in thin film configuration where the thickness is of the same order than the magnetic penetration length $\lambda$ is needed. At high magnetic fields, a recent theoretical work\cite{Rosen04} has successfully fitted the experimental curves of Figure 1 with a modified time-dependent Ginzburg Landau equation. In summary, dissipative phenomena governing the FF regime are very different between ``clean'' and ``dirty'' films although a clear linear $E(J)$ regime is observed in both cases.\\ 
%
%				FIGURE 4
Looking now at higher dissipation, properties of the jump into the normal state are discussed within the BS model for which the finite thermal exchange $h$ between the superconducting film and the substrate plays a role. Values of critical coordinates $(J^{*},E^{*})$ are confronted with the scaling law defined by equation $(6)$. First, the field $B_{t}$ is directly deduced from the $z(b_{t})$ dependence of the dissipative power $P^{*}=E^{*}\times J^{*}$. Inset of Figure \ref{figure5} shows good accordance between experimental data and theoretical curves with two ajustable parameters $B_{t}$ and $P^{*}_{LO}=E^{*}_{LO}\times J^{*}_{LO}$. In a second step, $E^{*}_{LO}$ and $J^{*}_{LO}$ are extracted from the $B_{t}$ (i.e. $z(b_{t})$) dependance of $J^{*}$ and $E^{*}$ according to the BS model. The curves  $J^{*}(z)$ and  $E^{*}(z)$ follow the dependence predicted by equations 33 and 35 of reference\cite{BS92} with only one adjustable parameter per equation, respectively $J^{*}_{LO}$ and $E^{*}_{LO}$. This analysis is performed in a large range of $T$ and $B_{a}$ as seen in Figure \ref{figure5}, which displays the expected scaling law predicted by equation \ref{eq6}. The extracted values of $B_{t}$ are about $5-10mT$ in our experimental  range for all samples, which corresponds to vortex densities around $2$ to $5$ $vortex/\mu m^{2}$. These low densities tend to show that the regime of quasiparticles  heating occurs in a large field domain even for low temperatures.
%
%				FIGURE 5
The field dependence  of $v^{*}$ in the form $B^{n}$ in both superconducting limits (Figure \ref{figure6}), which is not predicted in the frame of the LO model, gives another proof of the quasiparticles  heating mechanism in our samples. The experimental $n$ values are close to 0.4, which is in a reasonable agreement with the BS model, although some refinements of the theory are needed here. 
The highest velocities $v^{*}$ are found in the dirty limit.  For ``dirty'' films, the theoretical relation fits exactly the experimental $v^{*}(T)$ variation  (Inset of Figure \ref{figure6}) in the range $B_{a}>B_{t}$ by taking a constant heat exchange coefficient $h$ versus temperature. In contrast, $v^{*}(T)$ for ``clean'' niobium shows that $h$ will vary a lot with temperature. \\
Taking into account the good accordance between experimental data and theory, $\tau_{qp}(T)$ is extracted from the product $P^{*}_{LO}=J^{*}_{LO}\times E^{*}_{LO}$ and the normal conductivity $\sigma_{n}$ (equations 34 and 36 of reference\cite{BS92} giving $\tau_{qp}=2.67\frac{B_t}{P^{*}_{LO}}\frac{\sigma_n}{e}k_BT_c(1-t)$). This $\tau_{qp}$ calculation has the advantage to not depend on Fermi velocity $v_{F}$ which is difficult to know precisely. 
%
%				FIGURE 6
Two very different microscopic mechanics are revealed with the  $\tau_{qp}(T)$dependence for both superconducting limits. For ``dirty'' niobium films, quasiparticles  lifetime doesn't change much with temperature  and lays in the range $30-50$ $ps$, this order of magnitude being in accordance with data obtained from magneto-conductance measurements\cite{Hi90} and photonic $Nb$ detectors\cite{Sr96} (figure\ref{figure7}). \\
This order of magnitude is also compatible with a simple calculation considering that vortex motion is characterised by pair breaking upstream from the core and by quasiparticles  recombinaison downstream. Quasiparticles  inside the core can relax their extra energy given by the electric field only if their characteristic time of presence in this normal region is larger than $\tau_{qp}$. When the vortex motion becomes too fast, this energy relaxation can not occur anymore and an effective increase in quasiparticle  energy that will ultimately lead to the instability, takes place. In this description, $\tau_{qp}$ obeys the simple law 
$\xi/v_{max}=\tau_{qp}$. As $\xi \approx 10nm$ in the dirty sample at 7.8K, we obtain with $v_{max}=480 m/s$ (see Figure 3b), a characteristic time $\tau_{qp}=6.10^{-11}s$, in perfect agreement with the value obtained from the BS model. \\
In the clean sample the same analysis yields $\tau_{qp}=5.10^{-10}s$ for this reduced temperature t=0.9, which is again perfectly compatible with our experimental data ($\xi_{T=7.5K}=58 nm, v_{max}=110m/s$). \\
The quasiparticle energy relaxation process is however ruled by different mechanisms in the two samples. When the mean free path is shorter than the coherence length (dirty sample), the energy relation occurs within the vortex core. In the opposite situation (clean sample), the quasiparticles  can relax their energy given by the electric field only after covering a distance within the core equivalent to several $\xi$. In the microscopic picture where the quasiparticle energy rise comes from Andreev reflections at the vortex core walls, one can see that an increase of the quasiparticles  energy can occur below $v_{max}$ in clean samples but it is only above $v_{max}$ that quasiparticles  can escape from the core, leading to its shrinkage. This efficient process of quasiparticles energy rise, even at low $v_{v}$ in clean samples,  is coherent with the fact that $v_{max}$ is here lower than in dirty systems. 

%
%				FIGURE 7
The above discussion has to be linked to the study of the $\tau_{qp}(T)$ dependance which is related to the physical mechanisms attached to this parameter. For example, quasiparticles  recombination to Cooper pairs with emission of phonons follows a law in $e^{2\Delta (T)/k_{B}T}$ with the temperature variation of the superconductive gap being $\Delta (T)\simeq \Delta (0)(1-t)^{0.5}$. Figure 7a shows that the $\tau_{qp}(T)$ variation in the clean sample conveniently fits with a recombination model\cite{Dot97}. The only adjustable parameter $\Delta (0)$ is found to be about $2 k_{B}T_{c}$, higher than the BCS value $\Delta (0)\simeq 1.76 k_{B}T_{c}$. Such a deviation from the weak coupling BCS limit has already been reported in the case of Nb thin films \cite{PR98}. On the contrary, a quasi-constant $\tau_{qp}(T)$ variation in dirty samples show that scattering mechanisms dominate and that quasiparticles  are ejected from the core at energies $\delta E>\!>\Delta(0)$,  as suggested by S.B. Kaplan et al.\cite{Kap76}. The predominance of scattering mechanisms is coherent with the fact that quasiparticles  relax their energy in the core on a distance $l_{free}<\xi$. The limiting mechanism of the instability is thus the inelastic scattering within the core for the dirty sample. Recombination process on the other hand describes well the fact that quasiparticles  scattering within the core is not the predominant phenomena which restrains the electronic leakage. 
The limiting mechanism will take place outside the core, which is coherent with a recombination mechanism. Although an extensive microscopic calculation would be needed to describe the physics of a vortex core in motion,  we can already see that the phenomenology of clean and dirty samples will be quite different and is indeed associated to distinct behaviours. \\
Knowing $\tau_{qp}$ and $B_t$, heat transfer coefficients $h$ are extracted from equation (4).  Values of $h$ are of the same order of magnitude (around $2.7$ to $4.7$ $W/cm^{2}K$) for thin films on Al$_{2}$O$_{3}$ substrates for both superconducting limits. These values are one to two order of magnitude lower than $h$ values for epitaxial $BiSrCaCuO$ films on $SrTiO_{3}$ substrates\cite{Xiao98a} deduced from the first experimental analysis based on BS model, or for epitaxial $YBaCuO$ films on saphire substrates\cite{Xiao98}, but are 10 to 100 times higher $h$ found for $YBaCuO$ monocristals\cite{Xiao01} only glued. Present values seem realistic although a higher heat transfer coefficient would have been expected in the case of the clean (oriented) thin film. We however find that $h$ is independent on the electronic nature of $Nb$ films and shows high values of thermal resistance between substrate and superconducting film. In this way, the thermal contact  is improved with the insertion of a metallic layer between substrate and $Nb$ film (Fig. 7b).\\
As a final remark, we can compare our results obtained in the present work to those reported in our previous paper\cite{Vil03} considering only a non thermal LO process. A significant differences is found for the clean samples where the $\tau_{qp}(T)$ variation extracted from the BS model now follows a law relative to a recombination process with a very reasonable $\Delta (0)$. 
%
%------------------------------------------------------------CONCLUSIONS
%
\section{Conclusion}
We have compared flux flow  properties in clean and dirty superconducting limits of niobium thin films. For ``clean'' films, dissipation is essentially due to currents crossing vortex core with a radius $\xi(T)$. On the contrary, FF properties is governed by thermal effects in the vicinity of the vortex core in ``dirty'' films.  We have identified a non-linear regime at high dissipation in the frame of a modified Larkin Ovchinnikov model that takes into account a quasiparticles  heating effect. Non-equilibrium lifetime of quasiparticles  $\tau_{qp}$ in ``clean'' films is interpreted as a recombination process. Values of $\tau_{qp}$  are about $10$ to $100$ times (few $10ps$) shorter in ``dirty'' niobium films and reveal different microscopic mechanism for quasiparticles  relaxation suggesting high performances for hot electrons bolometers. We hope that this study will motivate future experimental and theoretical works to understand in details FF properties and non-linear regime in ``dirty'' superconducting limit of thin films.

\section{Acknowledgements}
We are grateful to  A. Sulpice for experimental support, C.J. Van der Beek for useful discussions, and T. Crozes for elaboration of thin films.

%------------------------------------------------------------BIBLIOGRAPHY

\end{document}